# View-Based Modeling of Function Nets

Hans Grönniger[1], Jochen Hartmann[2], Holger Krahn[1],
Stefan Kriebel[2], and Bernhard Rumpe[1]

[1]Institut für Software Systems Engineering, TU Braunschweig, Germany
[2]BMW Group, München, Germany

## Abstract

*This paper presents an approach to model features and function nets of automotive systems comprehensively. In order to bridge the gap between feature requirements and function nets, we describe an approach to describe both using a SysML-based notation. If requirements on the automotive system are changed by several developers responsible for different features, it is important for developers to have a good overview and understanding of the functions affected. We show that this can be comprehensively modeled using so called "feature views". In order to validate these views against the complete function nets, consistency checks are provided.*

## 1. Introduction

The task of developing automotive embedded systems is complex since a large number of functions from different vehicle domains interact in many ways. To master this complexity, system descriptions are necessary that support the process of developing such systems and that bridge the gap between requirements of different features to an integrated function net architecture. However, problems with notations and tools proposed or used today often exist:

- Tools that provide full views of the system do not scale to a large amount of functions.

- Notations that have their roots in computer science are likely to be not accepted by users with a different professional background.

- Time-constraints make it economically not reasonable to establish a new tool in the development process if re-engineering or re-modeling of the whole system is required.

Our contribution to the problem of modeling complex embedded automotive systems is thus guided by the following goals:

- It should be possible to integrate the modeling notation and supporting tools seamlessly into existing processes with little overhead for developers.

- A comprehensible description of functions, their structure, behavior and interactions with other functions should be supported. In particular, interesting functional interrelations should be presentable in a way such that they are comprehensible for all developers involved.

In this paper, we focus on the structural issues and show how our view-based approach can be used for modeling logical architectures, also called function nets in [13].

The rest of the paper is structured as follows. In Section 2 the problem of modeling comprehensible function nets is detailed and the requirements for a solution are presented. Section 3 describes our approach to view-based modeling of function nets. Section 4 presents related work and Section 5 concludes the paper.

## 2. Problem Statement

We assume a development process in which requirements are captured mainly textually (e.g., in DOORS [12]) by different people who are responsible for certain features. We use the term *feature* to denote functionalities perceptible by customers (e.g. a braking system) as common in requirements engineering / domain analysis (e.g., [3]). The features are realized by several functions which cooperate to achieve the desired functionality. The functions themselves may be needed by multiple features within the automotive system. For instance, the logic of the braking system is used by the normal brake employed by the driver

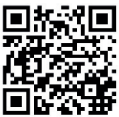

but also by safety or comfort systems like the adaptive cruise control.

The step from requirements to the actual realization of the system is a complex step that involves a lot of engineering work and coordination between different developers. Automotive architectures have been proposed to break down the complexity into manageable tasks on each architectural layer [2]. An example is shown in Fig. 1 in which requirements for all features are transformed into a function net that describes interacting logical functions that cooperatively fulfill all the requirements. In the software architecture, logical functions are aggregated or split into deployable units and detailed signal or function definitions, e.g., exact data types and value ranges are provided. On this architectural layer and below the AUTOSAR methodology can be applied [1]. The software architecture is then mapped to a technical architecture consisting of ECUs and busses. Software and technical architecture constitute the basis for a concrete realization of the final automotive system. In this paper, we concentrate on the transition from requirements to function nets.

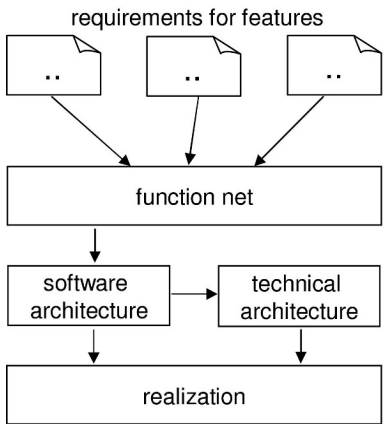

**Figure 1. Complex step from requirements to system realization**

The development is furthermore not being done from scratch but based on previous models, so a complete system from an earlier product line is already available. Typically, in the development of the next product line, changed requirements from various developers concerning new functions, function redesigns, or enhancements arise. However, many features that were present in the previous system will be reused (maybe in an enhanced version) in the next development cycle. This calls for a possibility to reuse functionality on the feature level.

The automotive system can be seen from two distinct viewpoints that both describe functions and their communication:

**Feature viewpoint.** A set of diagrams describes a single feature. One feature is usually treated separately from others in the requirements analysis phases and is realized by one or more functions.

**Logical viewpoint.** A set of diagrams describes a function and its hierarchical composition in form of an internal structure. The different functions may be realized on different ECUs and are therefore developed separately. Nevertheless, they must cooperate to achieve the common task.

When developing a notation that supports modeling these viewpoints comprehensibly, we have to keep in mind that the development process involves many people with different professional backgrounds (like computer science and engineering) that all need to be able to use the notation. Further, developing a modeling language should not be done from scratch but should be in line with existing standards and reuse ideas from other works.

Since there is no traceable connection from requirements to nets of logical functions in which the functionality of a feature is not explicitly conceivable, the development involves extensive re-engineering when the requirements change.

These observations lead us to the following problem statement:

a) Which notation is most appropriate, given the context and the intended use?

b) How can we model the feature requirements such that the effect of changes can easily be analyzed in the function net?

## 3. Proposed Solution

The demand for models that provide an overview of functions and their interactions on a more abstract level (logical architecture) than on the software and technical architecture level has also been stressed, e.g., in [13, 14]. We denote this logical architecture as function net models.

One standard technique to model complex systems in a comprehensible way is to use *hierarchical models*. This is certainly also appropriate for function nets since it allows us to model composite logical functions as a black-box and provide refinements of those functions that model their inner structure in more detail.

As explained above, especially the transition from the requirements to the function net is complex, because decisions about how a feature is realized and how distinct features use the same functions have to be made at the same time. Therefore we propose that it is useful to model each feature by a separate function net. These function nets explain how the feature can be realized. Then the functions of this *feature function net* can be related to the functions of the *automotive function net* describing the whole automotive system.

Modeling features as function nets that already represent parts of a possible logical architecture (w.r.t. notation and used concepts like functions and signals) helps to reduce the transition complexity because the conceptual distance between requirements and logical architecture is reduced. In addition the tasks of realizing the features as functions and the embedding of these functions into the whole automotive function net are now separate steps within the development process. This helps the developer to focus on a certain aspect at a time.

A complete function net model of the whole system may either be to complex to understand or be described on a too abstract level to be useful. We believe that modeling function nets for features is preferably done such that the model shows a complete definition of the logical feature of interest in addition with parts of other connected features on arbitrary hierarchy levels. Consequently, our solution supports *cross-hierarchy views* of function nets. Allowing arbitrary views of the system requires *consistency checks* that verify that the modeled view still conforms to the actual underlying system.

The different feature views on the system could be used to create the automotive function net by merging them. Techniques from requirements engineering research (e.g. [10]) might be used to create the automotive function net. The main problem in these approaches is that the views evolve during the design process and the merging process has to be re-applied partially, obeying prior results. We also doubt that theses approaches scale to large system and decided therefore against automatic function net merging. Our approach is restricted to checking for consistency but helps the developer to detect inconsistencies. The resolution has to be applied manually.

As the complexity of automotive systems steadily increases, it is no longer economically reasonable to develop such systems from scratch. The reuse of software is enabled by standardizing the software architecture by the AUTOSAR consortium [1] and therefore enables the development of reusable software components. But the effort in the development of an automotive system does not rely at most on the implementation part but also on developing the requirements. Therefore the proposed method allows the development of the requirements of the feature separately and also their modular redesign or substitution when shifting to the next car generation. The attached functions nets describing a single feature can then be checked against the evolved automotive function net. The approach is illustrated in Figure 2.

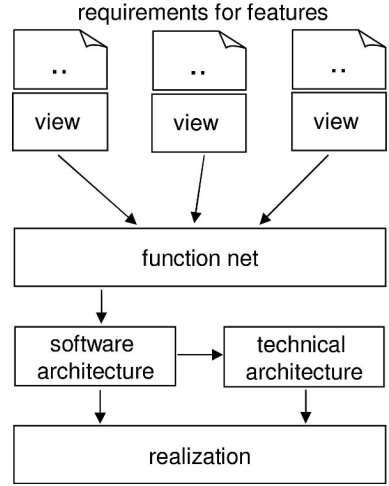

**Figure 2. Intermediate views to simplify transition from requirements to function nets**

### 3.1 SysML block diagrams

In [8] one of the authors investigated the use of UML and enhancements as an architecture description language. One of the results was that especially a hierarchical component-based notation was missing at that time. In [9] the use of UML-RT [11] is investigated for embedded real-time systems in general, whereas other previous work [13] has shown that function nets can be modeled with UML-RT. It is outlined that UML 2.0 [5] and SysML [6] are good candidates for substituting UML-RT.

Therefore we investigated among other notations the suitability of UML 2.0 and SysML for function net modeling. Our investigation showed that SysML block diagrams can be favored over UML composite structures, because they allow a more concise representation of systems. The detailed reasons are the following:

- SysML requires no strict two layered modeling like in the UML where each structured class consists of parts that in turn have no internal structure.

- SysML block diagrams allow modeling communication across multiple hierarchy layers without

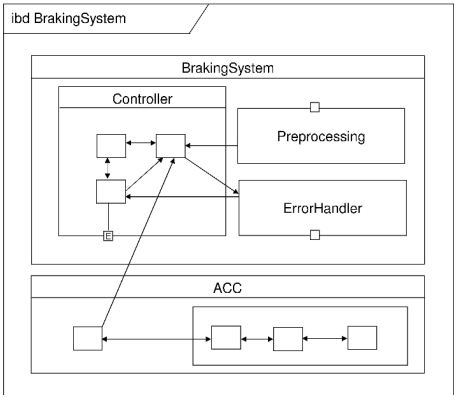

**Figure 3. Excerpt from an automotive function net (Signals are omitted to enhance the readability)**

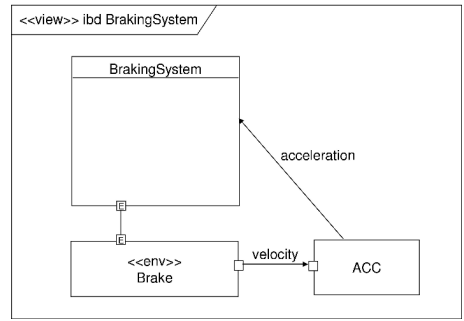

**Figure 4. View of the automotive function net describing the braking system and its environment**

the explicit use of port delegation.

- A SysML block abstracts from the strict instance/type division of the UML which complicates modeling architectures effectively.

- SysML distinguishes between the form of a diagram and its use. This was extremely helpful (as later shown) when we wanted to use the same diagram type with a different semantics.

The described port-delegation and the strict instance/type division originate from the modeling of object-oriented software, where each class is a point of variation. In modeling logical architectures this is not always the case as blocks might also be used to group subsystems which are not meant for separate reuse. SysML allows us to introduce types and therefore a reuse of subsystems where needed, in contrast to the UML which assumes constant reuse in all cases.

SysML block diagrams can be used by modeling functions of an automotive system as blocks. These blocks can be hierarchical decomposed into subblocks that define the internal structure. Blocks can be connected to each other via directed connectors that represent a communication relationship. The connector can be used across the block hierarchy but also ports can be used to describe a well-defined interface. To increase the reuse, blocks can optionally have a type that allows the multiple instantiation of a single block within a diagram. Figure 3 shows an example of such a diagram. Please note that for space reasons it is not complete in the sense that it contains all possible information nor describes a representative subset of an automotive subsystem.

Block diagrams can be used to model a complete automotive system. For organizational reasons the diagram can be split such that many orthogonal diagrams exist which describe the system in a readable size. Additional information for the blocks and signals can be stored efficiently in a database to allow queries about the stored information.

### 3.2 SysML block diagrams for feature views

The block diagrams can also be used to describe the functions needed to realize a feature of the system. In contrast to the already mentioned hierarchical modeling of the automotive system, blocks may occur in multiple diagrams. In addition to the already described elements of the SysML block diagram we provide extensions for modeling the physical environment of the electronic system. In discussions with developers of such systems it turned up that it is extremely helpful for the understanding of the system to include additional elements. By this approach complete closed loop controllers can be modeled instead of just considering the control part. We represented surrounding elements as ordinary blocks (marked by a special stereotype <<env>>) and non-signal communication by ports with a stereotype stating the type of communication like electric or hydraulic. Figure 4 shows a block diagram that represents a view of the diagram shown in Figure 3. The same elements like in the block diagram occur and elements of the environment are added to simplify the understanding of the feature.

The two described notations can be checked for consistency. The detailed relation between block diagrams and views is given by the following list of context conditions that must hold:

- Each block in a view not marked with a stereo-

type `<<env>>` must be part of the logical architecture in the block diagram.

- A hierarchy indicated in a view must be present in the logical architecture (although intermediate blocks may be left out).

- Communication relationships shown in a view must be present in the logical architecture. If the view indicates that certain signals are involved in a communication they must be stated in the architecture. If no signal is attached to a communication link in a view at least one signal must be present in the architecture. A communication relationship needs not be drawn to the exact target, also any superblock is sufficient.

## 4. Related Work

In [13] function net modeling with the UML-RT is described. We extended this approach by using the SysML for modeling function nets and explained its advantages. We supplement the approach by views that simplify the transition from requirements to early design phases.

In [10] view merging in the presence of incompleteness and inconsistency is described. The merging algorithm also simplifies the transition from requirements to early design phases like our approach. Especially the constant evolution of requirements during the development makes it difficult to apply such algorithms to our problem.

In [7, 15] service oriented modeling of automotive systems is explained. The service layer is similar to the modeling of features. In addition we explored how services can benefit from modeling the environment together with the feature.

In [4] the use of rich components is explained that employ a complex interface description including non-functional characteristics. In contrast to our approach rich components focus less on the seamless transition from requirements to function nets but assume an established predefined partitioning in components.

The AUTOSAR consortium [1] standardizes the software architecture of automotive system and allows the development of interchangeable software components. One main problem of this approach is that software architectures are too detailed in early development phases where functions nets are commonly accepted by developers.

## 5. Conclusion

In this paper we propose an approach to use block diagrams as provided by SysML to model individual vehicle functionalities, so called "features" of an automotive system and complete function nets using a similar notation. This similarity simplifies the seamless transition from stating the requirements to designing the system and reduces the necessary effort for feedback loops in the development cycle. It also allows switching viewpoints between feature-oriented requirements and function net architecture more easily.

A drawback of our approach might be the introduction of an additional modeling layer for views. While the results of smaller case-studies are promising, a detailed evaluation of the method with an example of realistic size still needs to be carried out.

The approach described in this paper focuses on structural aspects. We already identified timing properties and other physical requirements and constraints that will be annotated to both the function nets and their according views. Timing constraints are a special form of requirements that shall be expressed in the feature diagrams. For the function net the actual execution times of an implementation may be derived by formal analysis or measuring representative runs. In the future we will further explore consistency checks that can be derived from these two types of timing properties.